\documentclass{article}
\usepackage[authoryear,round]{natbib}

\title{A Tale of Two Referees}
\author{Hannes Leeb}
\date{\today}

\begin{document}
\sloppy
\setlength{\parskip}{1.5 \medskipamount}

\maketitle

\begin{abstract}
Success in academia hinges on publishing in top tier journals.
This requires innovative results. And this requires
clear and convincing presentation of said results.
Presentation can make the difference 
of one tier in journal level.
A lot of useful advice on this topic is available online
from well-respected outlets; see, for example,
\cite{ElOmar}; 
\cite{Gould}; 
\cite{Neiles}; 
\cite{Notz}; or
\cite{Sachdeva}. 
This text provides a different angle.
\end{abstract}

\section*{Two types of referee}

When your research is completed, it comes to preparing a manuscript for
submission. You select a target journal, study its style and
write your manuscript accordingly. In doing so, it is helpful to
look at what will happen after submission.

The handling editor will select a number of people to referee the
paper. Ideally, the editor picks at least one person whose specialty
is the subject area of your paper, an expert or `maven;'
the maven's opinion
on technical correctness and innovation will carry a
heavy weight.  And the editor also picks  someone to evaluate your 
paper's merits from a broader perspective, who is not myopically focused
on the particular subject area, a non-expert or `muggle.'\footnote{
	The muggle will almost certainly be an expert in his own area;
	but as far a your paper is concerned, he is a muggle.
}
Sometimes, there will be several or combinations of these two types;
sometimes, one of them will be missing.
As this is unknown to you at the time of writing, it is wise to 
assume that both a maven and a muggle will referee the paper.

You need to prepare the manuscript so as to get a positive report from
both the maven and the
muggle. In doing so, it is best to assume the worst of them. 
This facilitates a more detailed description of these two types of referee.

\section*{The moody muggle}

Try to look at things from that referee's perspective: He has to spend
several hours evaluating someone else's research in an area he is only vaguely
familiar with, hours that he would gladly spend otherwise.
Standard arguments in your area may not be known to him,
he will at best skim over the technical parts.  
His attention span will be short. 
And as soon as he has lost attention, he will write his report.
He is a moody muggle.
That said, the moody muggle rarely is openly hostile, just moody.
Your paper should be written so as to elicit a strong positive report 
from him.

You have about 5 pages to win over the moody muggle. For this, you need
to get him positively excited about your results and about their broader impact
on the field, without over-selling.  
Do this as non-technically as possible. Avoid formulae
as much as you can. Avoid insider talk. 
Accomplishing this in about 5 pages is crucial, but can be extremely difficult. 
The rest of the paper should be written so that the moody muggle
can get a superficial understanding by skimming through it.
The moody muggle is a smart guy. And if you can hold his attention long
enough to hammer home your main points and innovations, without 
getting him confused, you are halfway done.

\section*{The mean maven}

If the handling editor is any good, the maven will not be your friend.
He might be in competition with you, or your results might extend,
replace or contradict his own work. He might even have an
interest in delaying or avoiding publication of your paper.
Ergo, the mean maven.
You will not get an enthusiastically positive report from him.
The best you can hope for is a grudging admission that the results
are correct and innovative.

The mean maven does not need convincing that your paper's area is important;
he is working on the same stuff.
He does, however, have an interest that the journal is not polluted
by publication of
faulty, incomplete, irrelevant, incompetent or otherwise mediocre
work. You need to write your paper so you do not give him
any angle of attack.
The mean maven will check whether all the relevant literature is cited
(most likely, his name will be in the bibliography).
He will be vain and he will not respond well if you denigrate his own work. 
He will look for errors, clumsy arguments, possible generalizations,
or conceptual issues in your results.
The first two must be avoided at all costs. The other two
should be addressed in the paper. It is always better to discuss
an issue before the mean maven
can raise it in his report.

The good thing about the mean maven is that his attention span
is practically unlimited.
After the first 5 pages, the rest of your paper
should be focused on not giving him an opening.

\section*{Satisfying both}

Eliciting a positive report from both the moody muggle and the
mean maven can be conflicting objectives. As a global optimum may not
exist, an iterative approach is in order:
Write a first version of the manuscript
to satisfy the moody muggle. Then read this from the mean maven's perspective
and re-write the manuscript accordingly. 
Next, read from the moody muggle's perspective and re-write again.
Repeat this. Several times. Stop only when a complete pass-through from
both viewpoints does not lead to further changes.
Finally, submit and wait.
If you are very lucky, you can skip the next section.

\section*{Dealing with reports}

Critical reports will happen. The most infuriating ones are
one-line rejections. The most useful ones are detailed discussions.
Before deciding what to do next, wait until you can read the reports
with a sufficiently clear head.
Useful feedback is not necessarily pleasant feedback.
And even though some reports are positively inane, you should still
try to get the most out of them.
Very smart people have read your work and provided their feedback, however
detailed, however positive, however reasonable. 
Use it and revise the paper accordingly, irrespective of whether or
not it was rejected.

If the editor invites a revision, the replies to the referees should
be prepared with great care. Gauge who wrote a report, a muggle or a maven.
Muggles sometimes provide interesting feedback and suggestions,
not all of which are practical. If there was a mis-understanding,
eliminate its source and respond kindly.
A maven's suggestions should be followed to the letter, to the extent
this is possible. 
If this is not so,
respectfully but firmly stand your ground.

After a rejection, aspirations sometimes have to be adjusted.
Otherwise, just move on
to the next journal on your list.

{
\bibliographystyle{plainnat}
\bibliography{article}{}

\begin{thebibliography}{5}
\providecommand{\natexlab}[1]{#1}
\providecommand{\url}[1]{\texttt{#1}}
\expandafter\ifx\csname urlstyle\endcsname\relax
  \providecommand{\doi}[1]{doi: #1}\else
  \providecommand{\doi}{doi: \begingroup \urlstyle{rm}\Url}\fi

\bibitem[El-Omar(2014)]{ElOmar}
E.M. El-Omar.
\newblock How to publish a scientific manuscript in a high-impact journal.
\newblock \emph{Advances in Digestive Medicine}, {\bf 1}:\penalty0 105--109,
  2014.

\bibitem[Gould(2014)]{Gould}
J.~Gould.
\newblock How to get published in high-impact journals: An essential guide.
\newblock
  http://blogs.nature.com/naturejobs/2014/06/06/how-to-get-published-in-high-impact-journals-an-essential-guide/,
  2014.

\bibitem[Neiles et~al.(2015)Neiles, Carey, Araujo, Burkhart, Kirschman,
  LaBumbard, LaGrange, Maine, Rombenso, Wood, and Boyles]{Neiles}
B.~Neiles, C.S. Carey, A.~Araujo, D.~Burkhart, L.J. Kirschman, B.~LaBumbard,
  S.~LaGrange, J.J. Maine, A.M. Rombenso, M.N. Wood, and J.G. Boyles.
\newblock Writing your way into high impact factor journals.
\newblock \emph{Bulletin of the Ecological Society of America}, {\bf
  96}:\penalty0 312--316, 2015.

\bibitem[Notz and Kafadar(2011)]{Notz}
W.I. Notz and K.~Kafadar.
\newblock Tips on publishing and reviewing papers in statistics journals.
\newblock https://stattrak.amstat.org/2011/12/01/journaltips/, 2011.

\bibitem[Sachdeva(2020)]{Sachdeva}
S.M. Sachdeva.
\newblock 5 tips for publishing in a high impact journal.
\newblock
  https://www.elsevier.com/connect/5-tips-for-publishing-in-a-high-impact-journal,
  2020.

\end{thebibliography}
}

\end{document}